\documentclass[a4paper,aps,prd,10pt,preprintnumbers,showpacs,twocolumn,superscriptaddress,nofootinbib,amsmath,amssymb]{revtex4-2}
\usepackage{graphicx}
\usepackage[utf8]{inputenc}
\usepackage[T1]{fontenc}
\usepackage{cmap}
\usepackage{soul}
\usepackage{amsmath}

\begin{document}

\title{Late time decay of scalar and Dirac fields around an asymptotically de Sitter black hole in the Euler-Heisenberg electrodynamics}

\author{S. V. Bolokhov}\email{bolokhov-sv@rudn.ru}
\affiliation{Peoples' Friendship University of Russia (RUDN University), 6 Miklukho-Maklaya Street, Moscow, 117198, Russia}

\begin{abstract}
We compute the quasinormal modes of massive scalar and Dirac fields within the framework of asymptotically de Sitter black holes in Euler-Heisenberg non-linear electrodynamics. We pay particular attention to the regime $\mu M/m_{P}^2 \gg 1$, where $\mu$ and $M$ denote the masses of the field and the black hole, respectively, and $m_{P}$ represents the Planck mass, covering a range from primordial to large astrophysical black holes. Through time-domain integration, we demonstrate that, contrary to the asymptotically flat case, the quasinormal modes also dictate the asymptotic decay of fields. Employing the 6th order WKB formula, we derive a precise analytic approximation for quasinormal modes in the regime $\mu M/m_{P}^2 \gg 1$ without resorting to expansion in terms of the inverse multipole number. This analytic expression takes on a concise form in the limit of linear electrodynamics, represented by the Reissner-Nordström black holes.
\end{abstract}

\maketitle

\section{Introduction}

The development of perturbations of massless fields around various asymptotically flat black holes can be categorized into three distinct stages: firstly, an initial outburst influenced by the initial conditions of the perturbations; secondly, damped oscillations controlled by the quasinormal modes; and finally, in the limit of asymptotically late times $t \rightarrow \infty$, a power-law decay \cite{Price:1971fb,Bicak}. 

The evolution of perturbations of massive fields with various spins in an asymptotically flat background has been explored in several works \cite{Koyama:2000hj,Koyama:2001qw,Koyama:2001qw,Moderski:2001tk,Jing:2004zb,Konoplya:2006gq,Seahra:2004fg}. These studies have revealed distinct behaviors: the tails exhibit oscillations with a power-law envelope. Regarding massless fields in an asymptotically de Sitter background, it has been demonstrated in \cite{Brady:1996za} that they decay exponentially at asymptotic times, a decay pattern attributed to a set of quasinormal modes \cite{Konoplya:2022xid}. Furthermore, massive fields in $D$-dimensional Schwarzschild-de Sitter and Reissner-Nordström backgrounds exhibit exponential decay at asymptotic times, which may or may not be oscillatory \cite{Konoplya:2024ptj,Correa:2024xki}. These decay patterns are associated with modes of the pure de Sitter space deformed by a black hole \cite{Lopez-Ortega:2012xvr,Lopez-Ortega:2007vlo}. Quasinormal modes of such massive fields in Schwarzschild-de Sitter and Reissner-Nordström backgrounds have been extensively studied, as evidenced by works such as \cite{Zhidenko:2003wq,Konoplya:2004uk,Jansen:2017oag,Aragon:2020teq,Konoplya:2014lha}, with particular attention paid to the anomalous decay rate in the regime $\mu M/m_{P}^2 \gg 1$ \cite{Gonzalez:2022upu,Fontana:2020syy} and Strong Cosmic Censorship \cite{Cardoso:2017soq,Konoplya:2022kld}. Mathematical aspects of asymptotic decay around black holes embedded in de Sitter space were addressed in \cite{Dyatlov:2010hq, Dyatlov:2011jd, Dyatlov:2011zz, Hintz:2016gwb}. Recent discussions have also focused on the potential observation of massive fields in experiments involving very long waves \cite{NANOGrav:2023gor}, as highlighted in \cite{Konoplya:2023fmh}. The motivations to study evolution of perturbations of massive fields are also connected to effective massive fields produced by the extra-dimensional scenarios \cite{Seahra:2004fg,Konoplya:2023fmh,Ishihara:2008re} or magnetic fields \cite{Konoplya:2007yy,Konoplya:2008hj,Kokkotas:2010zd}.

While the decay of massive fields in asymptotically de Sitter black holes within linear electrodynamics (as described by the Reissner-Nordström solution) has been relatively well-studied recently, there has been a lack of similar investigations in the realm of non-linear electrodynamics. Despite the existence of numerous exotic models of non-linear electrodynamics, often lacking a correct weak field limit, our focus lies on Euler-Heisenberg electrodynamics \cite{Heisenberg:1936nmg}, which is a component of the quantum electrodynamics framework and enables the description of photon-photon scattering effects \cite{Karplus:1950zz,dEnterria:2013zqi,TOTEM:2021zxa}.

The relationship between the quasinormal modes of a test massless scalar field and a massive charged one in the asymptotically flat black hole background of the Einstein-Euler-Heisenberg theory, and their connection to the null-geodesic/eikonal quasinormal modes correspondence \cite{Cardoso:2008bp}, has been discussed in previous works such as \cite{Breton:2016mqh,Breton:2021mju}. However, these discussions did not include actual calculations of quasinormal modes. Additionally, quasinormal modes of gravitational perturbations have been investigated in \cite{Nomura:2021efi}. Therefore, to the best of our knowledge, there have been no studies on the quasinormal modes of test fields within the framework of the Einstein-Euler-Heisenberg theory, neither in the presence of a cosmological constant nor in asymptotically flat spacetime.

In this study, we investigate the quasinormal modes of a massive scalar and massless Dirac fields within the background of static charged asymptotically de Sitter black holes, as discovered in \cite{Magos:2020ykt}, within the framework of the Einstein-Euler-Heisenberg theory, allowing for a (positive) cosmological constant.

We demonstrate that the decay at asymptotically late times is governed by these quasinormal modes. By employing an expansion in powers of the large parameter $\mu M/m_{P}^2$, we derive accurate analytical expressions for the quasinormal modes. This regime holds particular significance for our investigation, as it encompasses a wide range of black hole masses, from primordial to galactic scales, even if including the lightest particle of the Standard Model, the electron, with $\mu \approx 9.1 \times 10^{-31} kg$. Furthermore, the analytical formula is simplified into a compact form in the Reissner-Nordström limit. In addition, we study quasinormal modes of massless scalar and Dirac field and show that the oscillation frequencies and damping rates are greatly suppressed, once the cosmological constant is tuned on.

Our work is organized as follows. The basic information on the underlying theory, the black hole metric, the wave-like equations and effective potentials are given in sec. II. In sec. III we review the two methods used for finding quasinormal modes: the WKB method and time-domain integration. Sec. IV is devoted to calculations of quasinormal modes. In sec. V we derive the analytical approximate formula for quasinormal modes in the Reissnser-Nordström limit. Finally, in the conclusion we summarize the obtained results. 

\section{The wave equation, and effective potentials}

The $D=4$ action of general relativity  $\Lambda$ coupled to the nonlinear electrodynamics (NLED) \cite{Plebanski,Salazar} and allowing for a non-zero cosmological constant has the form
\begin{equation}
S=\frac{1}{4\pi}\int_{M^4} d^4x \sqrt{-g}\left[\frac{1}{4}(R-2\Lambda)-
\mathcal{L}(F,G) \right].\label{action}
\end{equation}
Here $g$ is the determinant of the metric tensor, $R$ is the Ricci scalar,
$\mathcal{L}(F,G)$ is the non-linear electrodynamics Lagrangian, $F=\frac{1}{4}F_{\mu \nu} F^{\mu \nu}$ and $G=\frac{1}{4}F_{\mu \nu}
{^*F^{\mu \nu}}$ where $F_{\mu\nu}$ is the electromagnetic field strength
tensor and ${^*F^{\mu \nu}}=\epsilon _{\mu\nu\sigma\rho} F^{\sigma\rho}
/(2\sqrt{-g})$ is its dual. 

The Lagrangian density of the Euler-Heisenberg electrodynamics \cite{EH} is
\begin{equation}
\mathcal{L}(F,G)=-F+\frac{a}{2}F^2+ \frac{7a}{8} G^2,\label{Lagrangian}
\end{equation}
where $a=8 \alpha^2/45 m^4$ is the so-called Euler-Heisenberg parameter, which is responsible for  the intensity od the non-linear electrodynamics, $\alpha$ is
the fine structure constant, $m$ is the electron mass. In units  $c=1= \hbar$,  the Euler-Heisenberg paramete is of the order of $\alpha / E_c^2$.
In the limit $a=0$ the theory is reduced to the Maxwell electrodynamics $\mathcal{L}(F)=-F$.

The static spherically symmetric charged black hole in the Einstein-Euler-Heisenberg electrodynamics is described by the metric \cite{Magos:2020ykt},
\begin{equation}\label{metric}
  ds^2=-f(r)dt^2+\frac{dr^2}{f(r)}+r^2(d\theta^2+\sin^2\theta d\phi^2),
\end{equation}
where, 
$$
\begin{array}{rcl}
f(r)&=&\displaystyle 1 - \frac{2 M}{r} +\frac{Q^2}{r^2} - \frac{a Q^{4}}{20 r^6},\\
\end{array}
$$
where $Q$ is the effective charge, $a$ is the coupling of the non-linear electrodynamics,  and $M$ is the mass.  From here and on we will measure all dimensional quantities in units of $M$, i.~e., we take $M=1$.

The scalar ($\Phi$) and Dirac ($\Upsilon$) fields  obey the following general-covariant equations:
\begin{subequations}\label{coveqs}
\begin{eqnarray}\label{KGg}
\frac{1}{\sqrt{-g}}\partial_\mu \left(\sqrt{-g}g^{\mu \nu}\partial_\nu\Phi\right)&=&0,
\\\label{covdirac}
\gamma^{\alpha} \left( \frac{\partial}{\partial x^{\alpha}} - \Gamma_{\alpha} \right) \Upsilon&=&0,
\end{eqnarray}
\end{subequations}
where $\gamma^{\alpha}$ are (noncommutative) gamma matrices, and $\Gamma_{\alpha}$ are spin connections.
After separation of variables equations (\ref{coveqs}) are reduced to the wavelike form \cite{Kokkotas:1999bd,Berti:2009kk,Konoplya:2011qq}:
\begin{equation}\label{wave-equation}
\frac{\partial^2 \Psi}{\partial r_*^2} - \frac{\partial^2 \Psi}{\partial t^2} - V(r) \Psi = 0,
\end{equation}
where the ``tortoise coordinate'' $r_*$ is defined as follows
\begin{equation}\label{tortoise}
dr_*\equiv\frac{dr}{f(r)}.
\end{equation}

Here the effective potential for the massive scalar field is
\begin{equation}\label{potentialScalar}
V(r) =f(r) \left(\frac{f'(r)}{r}+\frac{\ell  (\ell +1)}{r^2}+\mu ^2\right).
\end{equation}
where $\ell=0, 1, 2, \ldots$ are the multipole numbers.
While for the massless Dirac field there are two isospectral potentials,
\begin{equation}
V_{\pm}(r) = W^2\pm\frac{dW}{dr_*}, \quad W\equiv \left(\ell+\frac{1}{2}\right)\frac{\sqrt{f(r)}}{r},
\end{equation}
which can be transformed one into another by the Darboux transformation. For the massless Dirac field $\ell=1/2, 3/2,  \ldots$.
Therefore, we will study  quasinormal modes for only one of the effective potentials, namely, $V_{+}(r)$, because the accuracy of the WKB method is usually higher in this case.

\begin{figure}
\resizebox{\linewidth}{!}{\includegraphics{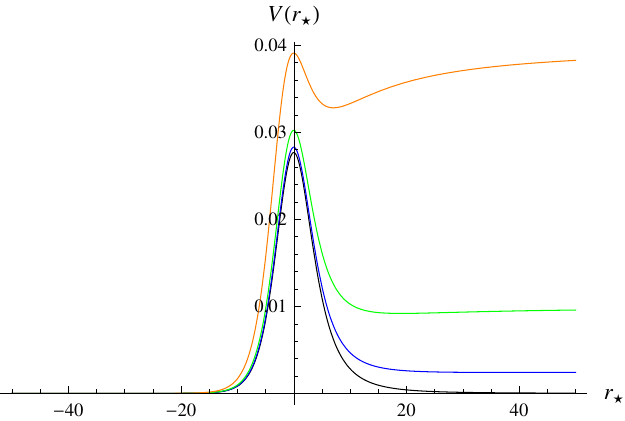}}
\caption{Potential as a function of the tortoise coordinate of the $\ell=0$ scalar field for the Magos-Breton black hole ($M=1$, $\Lambda=0$, $a=0.8$, $Q=0.5$): $\mu=0$ (black) $\mu=0.05$ (blue) $\mu=0.1$ (green) $\mu=0.2$ (orange).}\label{fig:PotScalar1}
\end{figure}

\begin{figure}
\resizebox{\linewidth}{!}{\includegraphics{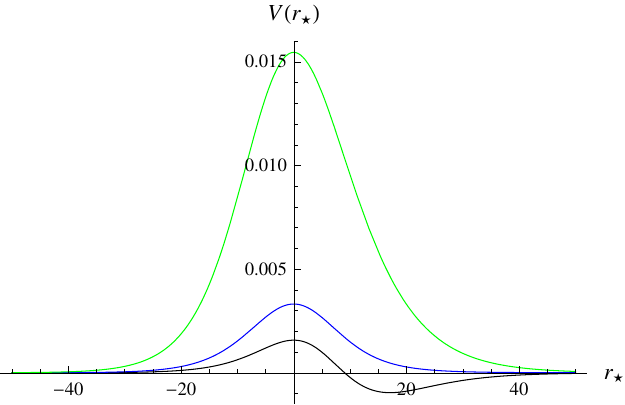}}
\caption{Potential as a function of the tortoise coordinate of the $\ell=0$ scalar field for the Magos-Breton black hole ($M=1$, $\Lambda=1/10$, $a=0.8$, $Q=0.5$): $\mu=0$ (black) $\mu=0.2$ (blue) $\mu=0.5$ (green).}\label{fig:PotScalar2}
\end{figure}

\begin{figure}
\resizebox{\linewidth}{!}{\includegraphics{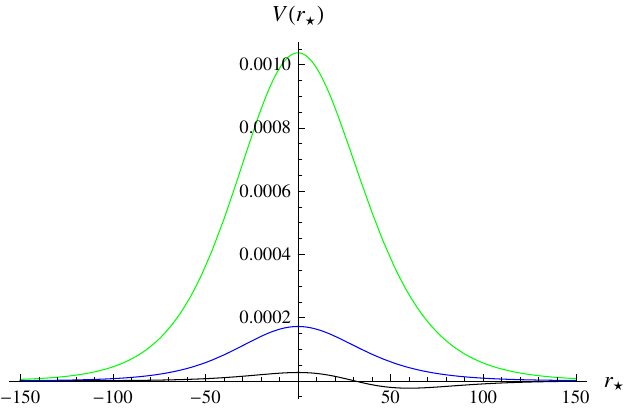}}
\caption{Potential as a function of the tortoise coordinate of the $\ell=0$ scalar field for the Magos-Breton black hole ($M=1$, $\Lambda=12/100$, $a=0.8$, $Q=0.5$): $\mu=0$ (black) $\mu=0.2$ (blue) $\mu=0.5$ (green) $\mu=1$.}\label{fig:PotScalar3}
\end{figure}

\section{The briefest summary of methods}

Quasinormal modes are complex values of $\omega= \omega_{Re} - i \omega_{Im}$, which correspond to solutions of the above second order differential master wave equation under specific boundary conditions: putely incoming wave at the event horizon and purely outgoing wave at the de Sitter horizon. Furthermore, $\omega_{Re}$ is the real oscillation frequency and $\omega_{Im}$ is proportional to the damping  rate.

\textbf{WKB}. The main method which will be used when searching for quasinormal modes in the frequency domain for large $\mu M$ is the WKB method \cite{Schutz:1985km,Iyer:1986np,Konoplya:2003ii}. It is based on the expansion of the solution of the wave equation in the  WKB series near the event horizon and de Sitter horizons and matching the asymptotic solutions with the Taylor expansion near the peak of the effective potential. 
Application of the WKB formula is effective here for  $\mu M \gg 1$ and $\Lambda \neq 0$, because the effective potential has a single maximum and decay monotonically towards both horizons. 
Furthermore we will use the Padé approximants  \cite{Matyjasek:2017psv} which greatly improves the accuracy. The WKB method with Padé approximants have been used in numerous recent works (see, for instance, \cite{Matyjasek:2021xfg,Konoplya:2023ahd,Malik:2023bxc,Paul:2023eep} and a review \cite{Konoplya:2019hlu}).
Comparison of the results obtained by the usual WKB method (even if used at higher orders) and the precise Leaver method, shows that without Padé approximants it is much less accurate, as for example, in \cite{Kodama:2009bf,Dubinsky:2024hmn}.

\textbf{Time-domain integration. } Evolution of perturbations in time-domain could be constructed for with the help of integration of the above wave-like equation at some fixed value of the radial coordinate. For this purpose, we use the null-cone variables $u = t - r_*$ and $v = t + r_*$ and the Gundlach-Price-Pullin discretization scheme \cite{Gundlach:1993tp}:
\begin{eqnarray}
\Psi(N) &=& \Psi(W) + \Psi(E) - \Psi(S) \nonumber \\
&& - \Delta^2V(S)\frac{\Psi(W) + \Psi(E)}{4} + \mathcal{O}(\Delta^4).\label{Discretization}
\end{eqnarray}
Here, the points are: $N \equiv (u + \Delta, v + \Delta)$, $W \equiv (u + \Delta, v)$, $E \equiv (u, v + \Delta)$, and $S \equiv (u, v)$. This method has also been used  in a number of works (see \cite{Konoplya:2020bxa, Bolokhov:2023dxq, Abdalla:2012si, Qian:2022kaq} and references therein) and showed excellent agreement of the fundamental mode with the precise Leaver method. The advantage of the time-domain integration method is that it works equally well for any values of $\mu M$.

%\textbf{Bernstein polynomial method.} If $\mu M \ll 1$, we will use the method based on the expansion of %the solution into Bernstein polynomials \cite{Fortuna:2020obg}. The details of this method and the {\it %Mathematica} code which we used can be found in \cite{Konoplya:2022zav}. 

\section{The late time decay and quasinormal modes}

%%%%%%%%%%%%%%%%%%%%%%%%%%%%%%%%%%%%%

\begin{table*}
\begin{tabular}{c c c c c c}
\hline
Q & $\Lambda$ & WKB6 Padé & WKB6 & difference & WKB3\\
\hline
$0$ & $0$ & $0.110792-0.104683 i$ & $0.110467-0.100816 i$ & $2.55\%$ & $0.104647-0.115197 i$\\
$0$ & $0.02$ & $0.098185-0.105302 i$ & $0.097287-0.101209 i$ & $2.91\%$ & $0.091823-0.112231 i$\\
$0$ & $0.04$ & $0.080126-0.102882 i$ & $0.079171-0.099742 i$ & $2.52\%$ & $0.075770-0.105903 i$\\
$0$ & $0.06$ & $0.059233-0.094709 i$ & $0.058298-0.092475 i$ & $2.17\%$ & $0.057178-0.094607 i$\\
$0$ & $0.08$ & $0.037652-0.080015 i$ & $0.036124-0.077461 i$ & $3.36\%$ & $0.036412-0.076878 i$\\
$0$ & $0.1$ & $0.019514-0.053922 i$ & $0.012614-0.049625 i$ & $14.2\%$ & $0.012986-0.047391 i$\\
$0$ & $0.11$ & $0.005573-0.005958 i$ & $0.000510-0.017183 i$ & $151.\%$ & $0.000616+0.015376 i$\\
$0.25$ & $0$ & $0.112025-0.104937 i$ & $0.111711-0.101093 i$ & $2.51\%$ & $0.105993-0.115237 i$\\
$0.25$ & $0.02$ & $0.099524-0.105579 i$ & $0.098640-0.101509 i$ & $2.87\%$ & $0.093217-0.112451 i$\\
$0.25$ & $0.04$ & $0.081743-0.103363 i$ & $0.080790-0.100195 i$ & $2.51\%$ & $0.077331-0.106440 i$\\
$0.25$ & $0.06$ & $0.061161-0.095625 i$ & $0.060226-0.093384 i$ & $2.14\%$ & $0.059007-0.095674 i$\\
$0.25$ & $0.08$ & $0.039815-0.081724 i$ & $0.038398-0.079278 i$ & $3.11\%$ & $0.038615-0.078863 i$\\
$0.25$ & $0.1$ & $0.020717-0.057893 i$ & $0.015333-0.053765 i$ & $11.0\%$ & $0.015787-0.051653 i$\\
$0.25$ & $0.11$ & $0.017394-0.021322 i$ & $0.003146-0.029033 i$ & $58.9\%$ & $0.002728-0.026699 i$\\
$0.5$ & $0$ & $0.116080-0.105642 i$ & $0.115852-0.101310 i$ & $2.76\%$ & $0.110666-0.115250 i$\\
$0.5$ & $0.02$ & $0.104010-0.106391 i$ & $0.103124-0.101843 i$ & $3.11\%$ & $0.097972-0.112989 i$\\
$0.5$ & $0.04$ & $0.087136-0.104737 i$ & $0.086060-0.101061 i$ & $2.81\%$ & $0.082536-0.107911 i$\\
$0.5$ & $0.06$ & $0.067522-0.098225 i$ & $0.066427-0.095681 i$ & $2.32\%$ & $0.064968-0.098703 i$\\
$0.5$ & $0.08$ & $0.046938-0.086549 i$ & $0.045642-0.084243 i$ & $2.69\%$ & $0.045650-0.084496 i$\\
$0.5$ & $0.1$ & $0.026913-0.067752 i$ & $0.023899-0.064471 i$ & $6.11\%$ & $0.024458-0.062860 i$\\
$0.5$ & $0.11$ & $0.019523-0.052933 i$ & $0.012498-0.048869 i$ & $14.4\%$ & $0.012835-0.046645 i$\\
$0.75$ & $0$ & $0.123027-0.104875 i$ & $0.118353-0.106245 i$ & $3.01\%$ & $0.122491-0.115146 i$\\
$0.75$ & $0.02$ & $0.111968-0.105809 i$ & $0.110366-0.101450 i$ & $3.01\%$ & $0.109592-0.113634 i$\\
$0.75$ & $0.04$ & $0.098573-0.106485 i$ & $0.096477-0.099037 i$ & $5.33\%$ & $0.094518-0.109954 i$\\
$0.75$ & $0.06$ & $0.081929-0.102215 i$ & $0.078888-0.095429 i$ & $5.68\%$ & $0.077819-0.103142 i$\\
$0.75$ & $0.08$ & $0.062934-0.093434 i$ & $0.059771-0.088685 i$ & $5.07\%$ & $0.059883-0.092801 i$\\
$0.75$ & $0.1$ & $0.042980-0.080541 i$ & $0.040021-0.077155 i$ & $4.93\%$ & $0.040816-0.078047 i$\\
$0.75$ & $0.11$ & $0.033160-0.071980 i$ & $0.029941-0.068763 i$ & $5.74\%$ & $0.030794-0.068335 i$\\
%$1.$ & $0$ & $0.128905-0.111925 i$ & $0.045331-2.158456 i$ & $1200.\%$ & $0.210285-0.153577 i$\\
%$1.$ & $0.02$ & $0.088820-0.123043 i$ & $0.748975+0.152001 i$ & $471.\%$ & $0.192210-0.149700 i$\\
%$1.$ & $0.04$ & $0.101391-0.106350 i$ & $1.098578+0.099252 i$ & $693.\%$ & $0.167307-0.139109 i$\\
%$1.$ & $0.06$ & $0.086322-0.105651 i$ & $0.846857+0.071626 i$ & $572.\%$ & $0.140502-0.126561 i$\\
%$1.$ & $0.08$ & $0.070614-0.108572 i$ & $0.548819+0.045197 i$ & $388.\%$ & $0.114071-0.113879 i$\\
%$1.$ & $0.1$ & $0.048770-0.112905 i$ & $0.313072+0.021094 i$ & $241.\%$ & $0.088908-0.101297 i$\\
%$1.$ & $0.11$ & $0.032380-0.116785 i$ & $0.222929+0.009696 i$ & $189.\%$ & $0.076888-0.094882 i$\\
\hline
\end{tabular}
\caption{Quasinormal modes of the $\ell=0$, $n=0$ scalar field for the Magos-Breton black hole calculated using the 6th  order WKB formulas with and without Padé approximants; $M=1$, $\mu=0$, $Q=0.5$, $a=10$.}
\label{tableI}
\end{table*}

\begin{table}
\begin{tabular}{c c c c c c}
\hline
Q & $\Lambda$ & WKB6 Padé & WKB6 & difference \\
\hline
$0$ & $0$ & $0.292930-0.097663 i$ & $0.292910-0.097762 i$ & $0.0327\%$\\
$0$ & $0.02$ & $0.260270-0.091013 i$ & $0.260265-0.091078 i$ & $0.0236\%$\\
$0$ & $0.04$ & $0.224677-0.082041 i$ & $0.224686-0.082088 i$ & $0.0200\%$\\
$0$ & $0.06$ & $0.185397-0.070067 i$ & $0.185417-0.070134 i$ & $0.0352\%$\\
$0$ & $0.08$ & $0.140400-0.054083 i$ & $0.140382-0.054171 i$ & $0.0595\%$\\
$0$ & $0.1$ & $0.081563-0.031217 i$ & $0.081560-0.031206 i$ & $0.0129\%$\\
$0$ & $0.11$ & $0.025490-0.009650 i$ & $0.025492-0.009651 i$ & $0.006\%$\\
$0.25$ & $0$ & $0.296092-0.097978 i$ & $0.296074-0.098071 i$ & $0.0304\%$\\
$0.25$ & $0.02$ & $0.263818-0.091458 i$ & $0.263813-0.091519 i$ & $0.0220\%$\\
$0.25$ & $0.04$ & $0.228721-0.082719 i$ & $0.228730-0.082763 i$ & $0.0184\%$\\
$0.25$ & $0.06$ & $0.190105-0.071132 i$ & $0.190125-0.071193 i$ & $0.0319\%$\\
$0.25$ & $0.08$ & $0.146153-0.055782 i$ & $0.146140-0.055872 i$ & $0.0583\%$\\
$0.25$ & $0.1$ & $0.090195-0.034305 i$ & $0.090191-0.034307 i$ & $0.0050\%$\\
$0.25$ & $0.11$ & $0.045354-0.017080 i$ & $0.045360-0.017067 i$ & $0.0298\%$\\
$0.5$ & $0$ & $0.306525-0.098879 i$ & $0.306502-0.098965 i$ & $0.0276\%$\\
$0.5$ & $0.02$ & $0.275463-0.092757 i$ & $0.275457-0.092815 i$ & $0.0197\%$\\
$0.5$ & $0.04$ & $0.241915-0.084723 i$ & $0.241920-0.084764 i$ & $0.0163\%$\\
$0.5$ & $0.06$ & $0.205324-0.074286 i$ & $0.205341-0.074335 i$ & $0.0237\%$\\
$0.5$ & $0.08$ & $0.164402-0.060767 i$ & $0.164397-0.060856 i$ & $0.0506\%$\\
$0.5$ & $0.1$ & $0.115290-0.042812 i$ & $0.115274-0.042838 i$ & $0.0249\%$\\
$0.5$ & $0.11$ & $0.083389-0.030753 i$ & $0.083387-0.030740 i$ & $0.0150\%$\\
$0.75$ & $0$ & $0.328324-0.100125 i$ & $0.328051-0.100437 i$ & $0.121\%$\\
$0.75$ & $0.02$ & $0.299516-0.094758 i$ & $0.299338-0.094923 i$ & $0.0773\%$\\
$0.75$ & $0.04$ & $0.268701-0.087940 i$ & $0.268658-0.088019 i$ & $0.0317\%$\\
$0.75$ & $0.06$ & $0.235662-0.079345 i$ & $0.235659-0.079425 i$ & $0.0323\%$\\
$0.75$ & $0.08$ & $0.199624-0.068703 i$ & $0.199622-0.068778 i$ & $0.0357\%$\\
$0.75$ & $0.1$ & $0.158946-0.055385 i$ & $0.158921-0.055439 i$ & $0.0356\%$\\
$0.75$ & $0.11$ & $0.135679-0.047333 i$ & $0.135657-0.047360 i$ & $0.0244\%$\\
$1.$ & $0$ & $0.378151-0.104427 i$ & $0.363906-0.115825 i$ & $4.65\%$\\
$1.$ & $0.02$ & $0.352828-0.099810 i$ & $0.343779-0.106933 i$ & $3.14\%$\\
$1.$ & $0.04$ & $0.326655-0.094422 i$ & $0.321208-0.098575 i$ & $2.01\%$\\
$1.$ & $0.06$ & $0.299138-0.088194 i$ & $0.296126-0.090387 i$ & $1.19\%$\\
$1.$ & $0.08$ & $0.270009-0.080958 i$ & $0.268517-0.081977 i$ & $0.641\%$\\
$1.$ & $0.1$ & $0.238847-0.072550 i$ & $0.238207-0.072947 i$ & $0.302\%$\\
$1.$ & $0.11$ & $0.222285-0.067841 i$ & $0.221892-0.068065 i$ & $0.195\%$\\
\hline
\end{tabular}
\caption{Quasinormal modes of the $\ell=1$, $n=0$ scalar field for the Magos-Breton black hole calculated using the 6th  order WKB formulas with and without Padé approximants; $M=1$, $\mu=0$, $Q=0.5$, $a=10$.}
\label{tableII}
\end{table}

\begin{table}
\begin{tabular}{c c c c c c}
\hline
Q & $\Lambda$ & WKB6 Padé & WKB6 & difference \\
\hline
$0$ & $0$ & $0.182643-0.096566 i$ & $0.182646-0.094935 i$ & $0.790\%$\\
$0$ & $0.02$ & $0.166921-0.086883 i$ & $0.166954-0.085473 i$ & $0.750\%$\\
$0$ & $0.04$ & $0.148957-0.076444 i$ & $0.149011-0.075250 i$ & $0.714\%$\\
$0$ & $0.06$ & $0.127601-0.064748 i$ & $0.127677-0.063816 i$ & $0.654\%$\\
$0$ & $0.08$ & $0.100547-0.050624 i$ & $0.100616-0.050066 i$ & $0.500\%$\\
$0$ & $0.1$ & $0.060612-0.030366 i$ & $0.060624-0.030245 i$ & $0.180\%$\\
$0$ & $0.11$ & $0.019236-0.009623 i$ & $0.019236-0.009617 i$ & $0.026\%$\\
$0.25$ & $0$ & $0.184745-0.096899 i$ & $0.184737-0.095328 i$ & $0.753\%$\\
$0.25$ & $0.02$ & $0.169170-0.087391 i$ & $0.169194-0.086031 i$ & $0.714\%$\\
$0.25$ & $0.04$ & $0.151445-0.077165 i$ & $0.151491-0.076006 i$ & $0.682\%$\\
$0.25$ & $0.06$ & $0.130504-0.065762 i$ & $0.130572-0.064842 i$ & $0.631\%$\\
$0.25$ & $0.08$ & $0.104262-0.052135 i$ & $0.104331-0.051558 i$ & $0.499\%$\\
$0.25$ & $0.1$ & $0.066740-0.033209 i$ & $0.066757-0.033052 i$ & $0.212\%$\\
$0.25$ & $0.11$ & $0.034071-0.016927 i$ & $0.034073-0.016908 i$ & $0.049\%$\\
$0.5$ & $0$ & $0.191805-0.097821 i$ & $0.191785-0.096301 i$ & $0.706\%$\\
$0.5$ & $0.02$ & $0.176660-0.088896 i$ & $0.176667-0.087588 i$ & $0.661\%$\\
$0.5$ & $0.04$ & $0.159637-0.079344 i$ & $0.159664-0.078232 i$ & $0.624\%$\\
$0.5$ & $0.06$ & $0.139900-0.068826 i$ & $0.139949-0.067922 i$ & $0.581\%$\\
$0.5$ & $0.08$ & $0.115926-0.056610 i$ & $0.115988-0.055981 i$ & $0.490\%$\\
$0.5$ & $0.1$ & $0.084127-0.040884 i$ & $0.084160-0.040615 i$ & $0.290\%$\\
$0.5$ & $0.11$ & $0.061779-0.029973 i$ & $0.061788-0.029873 i$ & $0.146\%$\\
$0.75$ & $0$ & $0.207017-0.099693 i$ & $0.208771-0.095085 i$ & $2.15\%$\\
$0.75$ & $0.02$ & $0.192612-0.091708 i$ & $0.193675-0.088361 i$ & $1.65\%$\\
$0.75$ & $0.04$ & $0.176769-0.083272 i$ & $0.177343-0.080892 i$ & $1.25\%$\\
$0.75$ & $0.06$ & $0.158923-0.074156 i$ & $0.159215-0.072540 i$ & $0.936\%$\\
$0.75$ & $0.08$ & $0.138237-0.064022 i$ & $0.138392-0.062999 i$ & $0.680\%$\\
$0.75$ & $0.1$ & $0.113144-0.052130 i$ & $0.113223-0.051578 i$ & $0.447\%$\\
$0.75$ & $0.11$ & $0.097914-0.045030 i$ & $0.097962-0.044681 i$ & $0.327\%$\\
$1.$ & $0$ & $0.243470-0.106079 i$ & $0.376830-0.017394 i$ & $60.3\%$\\
$1.$ & $0.02$ & $0.228967-0.099527 i$ & $0.351948-0.022854 i$ & $58.0\%$\\
$1.$ & $0.04$ & $0.214805-0.092395 i$ & $0.305854-0.032251 i$ & $46.7\%$\\
$1.$ & $0.06$ & $0.199783-0.084894 i$ & $0.257374-0.042324 i$ & $33.0\%$\\
$1.$ & $0.08$ & $0.183373-0.077046 i$ & $0.214701-0.050391 i$ & $20.7\%$\\
$1.$ & $0.1$ & $0.165046-0.068681 i$ & $0.179520-0.054276 i$ & $11.4\%$\\
$1.$ & $0.11$ & $0.154952-0.064212 i$ & $0.164110-0.054286 i$ & $8.05\%$\\
\hline
\end{tabular}
\caption{Quasinormal modes of the $\ell=1/2$, $n=0$ Dirac field for the Magos-Breton black hole calculated using the 6th  order WKB formulas with and without Padé approximants; $M=1$, $\mu=0$, $Q=0.5$, $a=10$.}
\label{tableIII}
\end{table}

\begin{table}
\begin{tabular}{c c c c c c}
\hline
Q & $\Lambda$ & WKB6 Padé & WKB6 & difference \\
\hline
$0$ & $0$ & $0.380041-0.096408 i$ & $0.380068-0.096366 i$ & $0.0128\%$\\
$0$ & $0.02$ & $0.344915-0.087145 i$ & $0.344931-0.087119 i$ & $0.00863\%$\\
$0$ & $0.04$ & $0.305421-0.076897 i$ & $0.305427-0.076882 i$ & $0.00497\%$\\
$0$ & $0.06$ & $0.259537-0.065157 i$ & $0.259538-0.065151 i$ & $0.0023\%$\\
$0$ & $0.08$ & $0.202957-0.050843 i$ & $0.202957-0.050842 i$ & $0.0005\%$\\
$0$ & $0.1$ & $0.121566-0.030408 i$ & $0.121566-0.030409 i$ & $0.0005\%$\\
$0$ & $0.11$ & $0.038485-0.009622 i$ & $0.038485-0.009622 i$ & $0\%$\\
$0.25$ & $0$ & $0.384179-0.096736 i$ & $0.384207-0.096691 i$ & $0.0134\%$\\
$0.25$ & $0.02$ & $0.349441-0.087647 i$ & $0.349458-0.087618 i$ & $0.00930\%$\\
$0.25$ & $0.04$ & $0.310507-0.077618 i$ & $0.310513-0.077601 i$ & $0.00560\%$\\
$0.25$ & $0.06$ & $0.265501-0.066183 i$ & $0.265503-0.066175 i$ & $0.00282\%$\\
$0.25$ & $0.08$ & $0.210551-0.052374 i$ & $0.210552-0.052372 i$ & $0.0008\%$\\
$0.25$ & $0.1$ & $0.133928-0.033265 i$ & $0.133928-0.033265 i$ & $0.0004\%$\\
$0.25$ & $0.11$ & $0.068204-0.016932 i$ & $0.068204-0.016932 i$ & $0.0002\%$\\
$0.5$ & $0$ & $0.397867-0.097669 i$ & $0.397903-0.097605 i$ & $0.0178\%$\\
$0.5$ & $0.02$ & $0.364345-0.089130 i$ & $0.364368-0.089087 i$ & $0.0130\%$\\
$0.5$ & $0.04$ & $0.327128-0.079782 i$ & $0.327139-0.079755 i$ & $0.00868\%$\\
$0.5$ & $0.06$ & $0.284744-0.069268 i$ & $0.284748-0.069254 i$ & $0.00509\%$\\
$0.5$ & $0.08$ & $0.234393-0.056905 i$ & $0.234394-0.056900 i$ & $0.0023\%$\\
$0.5$ & $0.1$ & $0.169095-0.040993 i$ & $0.169095-0.040993 i$ & $0.0003\%$\\
$0.5$ & $0.11$ & $0.123865-0.030013 i$ & $0.123866-0.030013 i$ & $0.0002\%$\\
$0.75$ & $0$ & $0.426584-0.098940 i$ & $0.426640-0.098813 i$ & $0.0316\%$\\
$0.75$ & $0.02$ & $0.395284-0.091427 i$ & $0.395326-0.091333 i$ & $0.0252\%$\\
$0.75$ & $0.04$ & $0.361101-0.083314 i$ & $0.361128-0.083249 i$ & $0.0190\%$\\
$0.75$ & $0.06$ & $0.323103-0.074389 i$ & $0.323117-0.074347 i$ & $0.0134\%$\\
$0.75$ & $0.08$ & $0.279727-0.064291 i$ & $0.279734-0.064268 i$ & $0.00830\%$\\
$0.75$ & $0.1$ & $0.227920-0.052313 i$ & $0.227922-0.052304 i$ & $0.0040\%$\\
$0.75$ & $0.11$ & $0.196826-0.045153 i$ & $0.196827-0.045149 i$ & $0.0022\%$\\
$1.$ & $0$ & $0.493604-0.101869 i$ & $0.491145-0.101338 i$ & $0.499\%$\\
$1.$ & $0.02$ & $0.465936-0.095459 i$ & $0.464368-0.095351 i$ & $0.331\%$\\
$1.$ & $0.04$ & $0.435949-0.089077 i$ & $0.435423-0.089110 i$ & $0.118\%$\\
$1.$ & $0.06$ & $0.404164-0.082674 i$ & $0.404062-0.082509 i$ & $0.0469\%$\\
$1.$ & $0.08$ & $0.369818-0.075611 i$ & $0.369828-0.075411 i$ & $0.0531\%$\\
$1.$ & $0.1$ & $0.331912-0.067784 i$ & $0.331947-0.067627 i$ & $0.0475\%$\\
$1.$ & $0.11$ & $0.311218-0.063519 i$ & $0.311251-0.063393 i$ & $0.0412\%$\\
\hline
\end{tabular}
\caption{Quasinormal modes of the $\ell=3/2$, $n=0$ Dirac field for the Magos-Breton black hole calculated using the 6th  order WKB formulas with and without Padé approximants; $M=1$, $\mu=0$, $Q=0.5$, $a=10$.}
\label{tableIV}
\end{table}

\begin{table*}
\begin{tabular}{c c c c c c}
\hline
$\mu$ & $\Lambda$ & WKB6 Padé & WKB6 & difference & WKB3\\
\hline
$0$ & $0$ & $0.116080-0.105642 i$ & $0.115852-0.101310 i$ & $2.76\%$ & $0.110666-0.115250 i$\\
$0$ & $0.06$ & $0.067522-0.098225 i$ & $0.066427-0.095681 i$ & $2.32\%$ & $0.064968-0.098703 i$\\
$0$ & $0.1$ & $0.026913-0.067752 i$ & $0.023899-0.064471 i$ & $6.11\%$ & $0.024458-0.062860 i$\\
$0$ & $0.11$ & $0.019523-0.052933 i$ & $0.012498-0.048869 i$ & $14.4\%$ & $0.012835-0.046645 i$\\
$0.1$ & $0$ & $0.119018-0.097798 i$ & $0.117662-0.093929 i$ & $2.66\%$ & $0.112512-0.108087 i$\\
$0.1$ & $0.06$ & $0.072445-0.094358 i$ & $0.071348-0.090179 i$ & $3.63\%$ & $0.068926-0.093952 i$\\
$0.1$ & $0.1$ & $0.035520-0.067744 i$ & $0.028351-0.060307 i$ & $13.5\%$ & $0.027894-0.060086 i$\\
$0.1$ & $0.11$ & $0.042042-0.044976 i$ & $0.017166-0.043863 i$ & $40.4\%$ & $0.015868-0.044416 i$\\
$0.2$ & $0$ & $0.135828-0.071872 i$ & $0.031923+0.359034 i$ & $288.\%$ & $0.120469-0.110278 i$\\
$0.2$ & $0.06$ & $0.085920-0.070817 i$ & $0.085952-0.068039 i$ & $2.49\%$ & $0.081326-0.075990 i$\\
$0.2$ & $0.1$ & $0.029534-0.044307 i$ & $0.035163-0.054037 i$ & $21.1\%$ & $0.038109-0.050141 i$\\
$0.2$ & $0.11$ & $0.018815-0.029700 i$ & $0.019358-0.042535 i$ & $36.5\%$ & $0.024373-0.036289 i$\\
%$0.6$ & $0$ &  &  &  & \\
$0.6$ & $0.06$ & $0.268545-0.054068 i$ & $0.267168-0.055491 i$ & $0.723\%$ & $0.269399-0.053741 i$\\
$0.6$ & $0.1$ & $0.143906-0.038361 i$ & $0.143896-0.038354 i$ & $0.0081\%$ & $0.144274-0.038470 i$\\
$0.6$ & $0.11$ & $0.103094-0.029018 i$ & $0.103093-0.029005 i$ & $0.0119\%$ & $0.103230-0.029077 i$\\
%$1.$ & $0$ &  &  &  & \\
$1.$ & $0.06$ & $0.449934-0.054157 i$ & $0.449935-0.054161 i$ & $0.00107\%$ & $0.449993-0.054129 i$\\
$1.$ & $0.1$ & $0.244446-0.038382 i$ & $0.244442-0.038384 i$ & $0.0016\%$ & $0.244498-0.038387 i$\\
$1.$ & $0.11$ & $0.175777-0.029026 i$ & $0.175777-0.029026 i$ & $0.\times 10^{\text{-4}}\%$ & $0.175802-0.029032 i$\\
%$5.$ & $0$ &  &  &  & \\
$5.$ & $0.06$ & $2.260669-0.054679 i$ & $2.260669-0.054679 i$ & $0\%$ & $2.260670-0.054679 i$\\
$5.$ & $0.1$ & $1.235401-0.038499 i$ & $1.235401-0.038499 i$ & $0\%$ & $1.235402-0.038499 i$\\
$5.$ & $0.11$ & $0.889826-0.029073 i$ & $0.889826-0.029073 i$ & $0\%$ & $0.889827-0.029073 i$\\
%$10.$ & $0$ &  &  &  & \\
$10.$ & $0.06$ & $4.522132-0.054696 i$ & $4.522132-0.054696 i$ & $0\%$ & $4.522132-0.054696 i$\\
$10.$ & $0.1$ & $2.471656-0.038504 i$ & $2.471656-0.038504 i$ & $0\%$ & $2.471656-0.038504 i$\\
$10.$ & $0.11$ & $1.780347-0.029075 i$ & $1.780347-0.029075 i$ & $0\%$ & $1.780347-0.029075 i$\\
\hline
\end{tabular}
\caption{Quasinormal modes of the $\ell=0$, $n=0$ scalar field for the Magos-Breton black hole calculated using the 6th  order WKB formulas with and without Padé approximants; $M=1$, $Q=0.5$, $a=10$.}
\label{tableV}
\end{table*}

\begin{table*}
\begin{tabular}{c c c c c c}
\hline
$\mu$ & $\Lambda$ & WKB6 Padé & WKB6 & difference & WKB3\\
\hline
$0$ & $0$ & $0.306525-0.098879 i$ & $0.306502-0.098965 i$ & $0.0276\%$ & $0.304905-0.099171 i$\\
$0$ & $0.06$ & $0.205324-0.074286 i$ & $0.205341-0.074335 i$ & $0.0237\%$ & $0.204832-0.074375 i$\\
$0$ & $0.1$ & $0.115290-0.042812 i$ & $0.115274-0.042838 i$ & $0.0249\%$ & $0.115222-0.042987 i$\\
$0$ & $0.11$ & $0.083389-0.030753 i$ & $0.083387-0.030740 i$ & $0.0150\%$ & $0.083269-0.030882 i$\\
$0.1$ & $0$ & $0.310716-0.096413 i$ & $0.310694-0.096510 i$ & $0.0307\%$ & $0.309079-0.096695 i$\\
$0.1$ & $0.06$ & $0.208989-0.073036 i$ & $0.208993-0.073095 i$ & $0.0266\%$ & $0.208430-0.073154 i$\\
$0.1$ & $0.1$ & $0.117770-0.042451 i$ & $0.117763-0.042467 i$ & $0.0145\%$ & $0.117648-0.042614 i$\\
$0.1$ & $0.11$ & $0.085235-0.030621 i$ & $0.085236-0.030604 i$ & $0.0184\%$ & $0.085103-0.030723 i$\\
$0.2$ & $0$ & $0.323379-0.088819 i$ & $0.323356-0.088917 i$ & $0.0300\%$ & $0.321694-0.089040 i$\\
$0.2$ & $0.06$ & $0.220154-0.069594 i$ & $0.220151-0.069624 i$ & $0.0132\%$ & $0.219338-0.069725 i$\\
$0.2$ & $0.1$ & $0.124987-0.041601 i$ & $0.124987-0.041603 i$ & $0.0017\%$ & $0.124768-0.041694 i$\\
$0.2$ & $0.11$ & $0.090542-0.030299 i$ & $0.090545-0.030290 i$ & $0.0094\%$ & $0.090417-0.030348 i$\\
%$0.6$ & $0$ &  &  &  & \\
$0.6$ & $0.06$ & $0.329426-0.056060 i$ & $0.329771-0.055674 i$ & $0.155\%$ & $0.329906-0.056276 i$\\
$0.6$ & $0.1$ & $0.186277-0.038879 i$ & $0.186285-0.038857 i$ & $0.0124\%$ & $0.186369-0.038915 i$\\
$0.6$ & $0.11$ & $0.134972-0.029229 i$ & $0.134973-0.029223 i$ & $0.0041\%$ & $0.135009-0.029246 i$\\
%$1.$ & $0$ &  &  &  & \\
$1.$ & $0.06$ & $0.486310-0.053902 i$ & $0.486277-0.053937 i$ & $0.00988\%$ & $0.486466-0.053894 i$\\
$1.$ & $0.1$ & $0.271074-0.038384 i$ & $0.271073-0.038384 i$ & $0.\times 10^{\text{-4}}\%$ & $0.271123-0.038393 i$\\
$1.$ & $0.11$ & $0.196003-0.029032 i$ & $0.196003-0.029031 i$ & $0.0001\%$ & $0.196023-0.029036 i$\\
%$5.$ & $0$ &  &  &  & \\
$5.$ & $0.06$ & $2.267674-0.054630 i$ & $2.267674-0.054630 i$ & $0\%$ & $2.267674-0.054630 i$\\
$5.$ & $0.1$ & $1.240822-0.038487 i$ & $1.240822-0.038487 i$ & $0\%$ & $1.240822-0.038487 i$\\
$5.$ & $0.11$ & $0.893992-0.029068 i$ & $0.893992-0.029068 i$ & $0\%$ & $0.893993-0.029068 i$\\
%$10.$ & $0$ &  &  &  & \\
$10.$ & $0.06$ & $4.525629-0.054684 i$ & $4.525629-0.054684 i$ & $0\%$ & $4.525629-0.054684 i$\\
$10.$ & $0.1$ & $2.474367-0.038501 i$ & $2.474367-0.038501 i$ & $0\%$ & $2.474367-0.038501 i$\\
$10.$ & $0.11$ & $1.782432-0.029074 i$ & $1.782432-0.029074 i$ & $0\%$ & $1.782432-0.029074 i$\\
\hline
\end{tabular}
\caption{Quasinormal modes of the $\ell=1$, $n=0$ scalar field for the Magos-Breton black hole calculated using the 6th  order WKB formulas with and without Padé approximants; $M=1$, $Q=0.5$, $a=10$.}
\label{tableVI}
\end{table*}

\begin{table}
\begin{tabular}{c c c c c c}
\hline
a & $\Lambda$ & WKB6 Padé & WKB6 & difference \\
\hline
$0$ & $0$ & $0.352583-0.097195 i$ & $0.352625-0.097208 i$ & $0.0119\%$\\
$0$ & $0.04$ & $0.298211-0.087100 i$ & $0.298226-0.087113 i$ & $0.0063\%$\\
$0$ & $0.06$ & $0.268561-0.080388 i$ & $0.268580-0.080406 i$ & $0.0093\%$\\
$0$ & $0.1$ & $0.202081-0.062623 i$ & $0.202080-0.062677 i$ & $0.0257\%$\\
$1.$ & $0$ & $0.352602-0.097619 i$ & $0.352567-0.097755 i$ & $0.0383\%$\\
$1.$ & $0.04$ & $0.298111-0.087476 i$ & $0.298107-0.087548 i$ & $0.0231\%$\\
$1.$ & $0.06$ & $0.268424-0.080713 i$ & $0.268425-0.080777 i$ & $0.0229\%$\\
$1.$ & $0.1$ & $0.201846-0.062841 i$ & $0.201837-0.062904 i$ & $0.0302\%$\\
$5.$ & $0$ & $0.352513-0.099254 i$ & $0.351979-0.100035 i$ & $0.258\%$\\
$5.$ & $0.04$ & $0.297747-0.088914 i$ & $0.297467-0.089244 i$ & $0.139\%$\\
$5.$ & $0.06$ & $0.267861-0.082002 i$ & $0.267706-0.082195 i$ & $0.0884\%$\\
$5.$ & $0.1$ & $0.200898-0.063657 i$ & $0.200853-0.063746 i$ & $0.0472\%$\\
$10.$ & $0$ & $0.352148-0.101128 i$ & $0.350512-0.102929 i$ & $0.664\%$\\
$10.$ & $0.04$ & $0.297110-0.090524 i$ & $0.296397-0.091197 i$ & $0.316\%$\\
$10.$ & $0.06$ & $0.267055-0.083424 i$ & $0.266669-0.083777 i$ & $0.187\%$\\
$10.$ & $0.1$ & $0.199706-0.064560 i$ & $0.199626-0.064656 i$ & $0.0597\%$\\
$15.$ & $0$ & $0.351606-0.102823 i$ & $0.348466-0.105640 i$ & $1.15\%$\\
$15.$ & $0.04$ & $0.296365-0.091971 i$ & $0.295188-0.092880 i$ & $0.479\%$\\
$15.$ & $0.06$ & $0.266179-0.084695 i$ & $0.265583-0.085121 i$ & $0.262\%$\\
$15.$ & $0.1$ & $0.198519-0.065350 i$ & $0.198424-0.065426 i$ & $0.0583\%$\\
\hline
\end{tabular}
\caption{Quasinormal modes of the $\ell=1$, $n=0$ scalar field for the Magos-Breton black hole calculated using the 6th  order WKB formulas with and without Padé approximants; $M=1$, $\mu =0$, $Q=0.9$.}
\label{tableVII}
\end{table}

\begin{table}
\begin{tabular}{c c c c c c}
\hline
a & $\Lambda$ & WKB6 Padé & WKB6 & difference \\
\hline
$0$ & $0.02$ & $1.364589-0.046519 i$ & $1.364589-0.046519 i$ & $1 \cdot 10^{-5}\%$ \\
$0$ & $0.04$ & $1.161257-0.055282 i$ & $1.161257-0.055282 i$ & $8 \cdot 10^{-6}\%$\\
$0$ & $0.06$ & $1.000119-0.057744 i$ & $1.000119-0.057744 i$ & $2 \cdot 10^{-5}\%$ \\
$0$ & $0.1$ & $0.713168-0.052307 i$ & $0.713168-0.052308 i$ & $4 \cdot 10^{-5}\%$\\
$1.$ & $0.02$ & $1.364587-0.046521 i$ & $1.364587-0.046520 i$ & $2 \cdot 10^{-5}\%$ \\
$1.$ & $0.04$ & $1.161244-0.055288 i$ & $1.161244-0.055288 i$ & $8 \cdot 10^{-6}\%$\\
$1.$ & $0.06$ & $1.000079-0.057758 i$ & $1.000079-0.057759 i$ & $2 \cdot 10^{-5}\%$ \\
$1.$ & $0.1$ & $0.712972-0.052342 i$ & $0.712972-0.052342 i$ & $3 \cdot 10^{-5}\%$\\
$5.$ & $0.02$ & $1.364577-0.046526 i$ & $1.364577-0.046526 i$ & $2 \cdot 10^{-5}\%$ \\
$5.$ & $0.04$ & $1.161190-0.055312 i$ & $1.161190-0.055312 i$ & $8 \cdot 10^{-6}\%$\\
$5.$ & $0.06$ & $0.999917-0.057816 i$ & $0.999917-0.057816 i$ & $2\cdot 10^{-5}\%$ \\
$5.$ & $0.1$ & $0.712189-0.052476 i$ & $0.712189-0.052476 i$ & $0\%$ \\
$10.$ & $0.02$ & $1.364565-0.046532 i$ & $1.364565-0.046532 i$ & $2 \cdot 10^{-5}\%$ \\
$10.$ & $0.04$ & $1.161122-0.055342 i$ & $1.161122-0.055342 i$ & $1 \cdot 10^{-5}\%$\\
$10.$ & $0.06$ & $0.999715-0.057887 i$ & $0.999714-0.057887 i$ & $4 \cdot 10^{-5}\%$\\
$10.$ & $0.1$ & $0.711214-0.052639 i$ & $0.711214-0.052639 i$ & $2 \cdot 10^{-5}\%$ \\
$15.$ & $0.02$ & $1.364552-0.046538 i$ & $1.364553-0.046539 i$ & $1 \cdot 10^{-5}\%$\\
$15.$ & $0.04$ & $1.161055-0.055372 i$ & $1.161055-0.055372 i$ & $2 \cdot 10^{-5}\%$ \\
$15.$ & $0.06$ & $0.999513-0.057957 i$ & $0.999512-0.057957 i$ & $3\cdot 10^{-5}\%$\\
$15.$ & $0.1$ & $0.710244-0.052795 i$ & $0.710244-0.052795 i$ & $6 \cdot 10^{-5}\%$\\
\hline
\end{tabular}
\caption{Quasinormal modes of the $\ell=0$, $n=0$ scalar field for the Magos-Breton black hole calculated using the 6th  order WKB formulas with and without Padé approximants; $M=1$, $\mu=2$, $Q=0.9$.}
\label{tableVIII}
\end{table}

\begin{figure}
\resizebox{\linewidth}{!}{\includegraphics{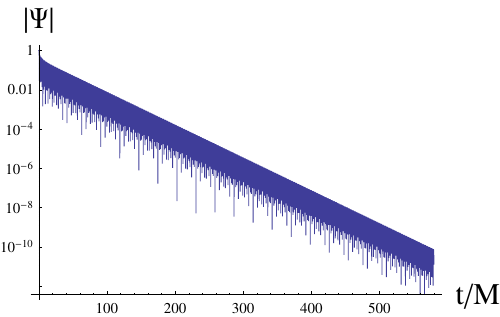}}
\caption{Time-domain profile for a massive scalar field in the background of the Magos-Breton black hole: $\mu M = 10$, $\Lambda M^2 = 0.1$, $Q=0.5$, $\ell=0$. The asymptotic decay is represented by the quasinormal frequency  $\omega M = 2.4728 - 0.0384857 i$. The 6th order WKB formula gives $\omega M = 2.472407 - 0.038504 i$.} \label{fig:timedomain1}
\end{figure}

\begin{figure}
\resizebox{\linewidth}{!}{\includegraphics{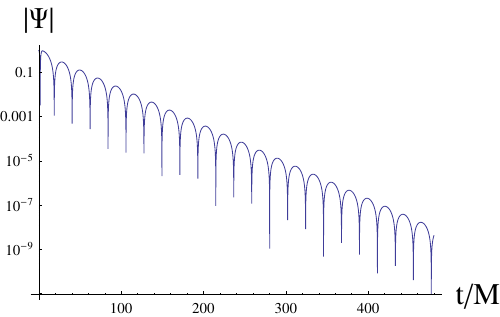}}
\caption{Time-domain profile for a massive scalar field in the background of the Magos-Breton black hole: $\mu M = 0.6$, $\Lambda M^2 = 0.1$, $Q=0.5$, $\ell=0$. The asymptotic decay is represented by the quasinormal frequency  $\omega M = 0.143994 - 0.0383527 i$. The 6th order WKB formula  gives $\omega M = 0.143951 - 0.038356 i$.} \label{fig:timedomain1}
\end{figure}

\begin{figure}
\resizebox{\linewidth}{!}{\includegraphics{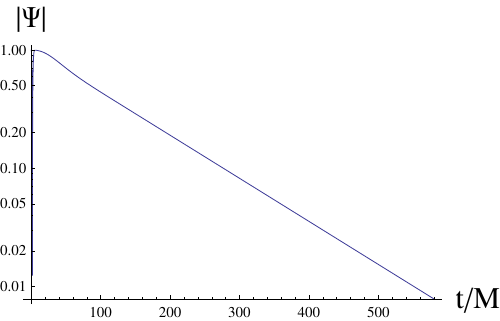}}
\caption{Time-domain profile for a massive scalar field in the background of the Magos-Breton black hole: $\mu M = 0.1$, $\Lambda M^2 = 0.1$, $Q=0.5$, $\ell=0$. The asymptotic decay is represented by the quasinormal frequency  $\omega M = - 0.0083962 i$.} \label{fig:timedomain1}
\end{figure}

\begin{figure*}
\resizebox{\linewidth}{!}{\includegraphics{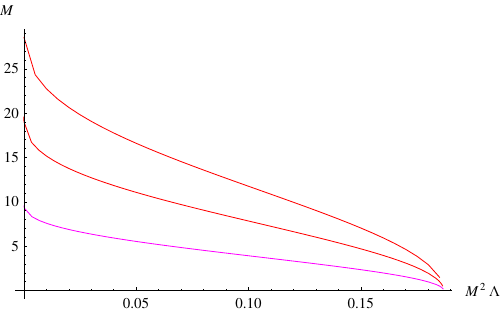}\includegraphics{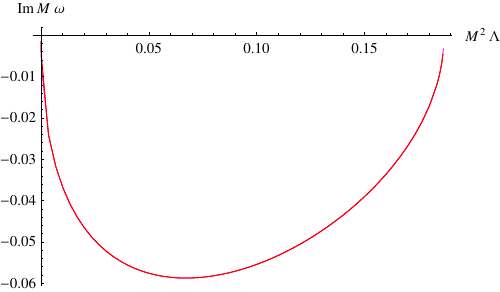}}
\caption{Real (left panel) and imaginary (right panel) parts of quasinormal frequencies governing the asymptotic decay at $\ell=0$ the Magos-Breton black hole, $Q=a=1$: $\mu M = 10$, $20$ and $30$ from bottom to top. The damping rate for different $\mu M$ is almost indistinguishable.}\label{fig:Q1}
\end{figure*}

\begin{figure}
\resizebox{\linewidth}{!}{\includegraphics{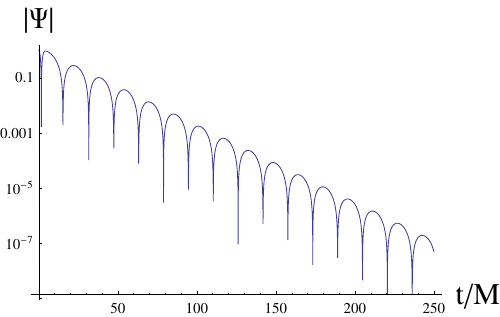}}
\caption{Time-domain profile for a massive scalar field in the background of the Magos-Breton black hole: $\mu = 0$, $\Lambda M^2 = 0.1$, $Q=0.9$, $\ell=1$. The asymptotic decay is represented by the quasinormal frequency  $\omega M = 0.199692 - 0.0645261 i$. The 6th order WKB formula in table \ref{tableVII} gives $\omega M = 0.199706 - 0.064560 i$.} \label{fig8}
\end{figure}

An important feature of asymptotically de Sitter black holes, which is appropriate to both the massless and massive cases, is the presence of two branches of modes. One branch could be called "black hole branch", because when the cosmological constant goes to zero, its frequencies transition to those for asymptotically flat black hole. In other words, these are modes of the black hole corrected by a non zero cosmological constant. The other branch consists of modes which transition into the modes of empty de Sitter space described in  in the limit of small black hole mass $M$. These modes are responsible for the asymptotic decay, as was shown for the Schrazschild-de Sitter case in \cite{Konoplya:2024ptj,Konoplya:2022xid}. 

Here we will consider two qualitatively different types of spectra: those for the massive and massless fields. 
Quasinormal modes of a massive scalar field in the asymptotically flat spacetime allow for an arbitrarily long lived quasinormal modes (quasiresonances) at some specific values of mass $\mu$ \cite{Ohashi:2004wr,Zhidenko:2006rs,Konoplya:2017tvu,Churilova:2019qph}. However, as was shown in \cite{Konoplya:2004wg} asymptotically de Sitter black holes have no such arbitrarily long lived modes. Here we see that while the cosmological constant suppresses the damping rate, the modes with very small damping rate, indicating the appearance of quasi-resonances, have not been observed, in concordance with \cite{Konoplya:2004wg}.

\subsection{Massless limit}

Quasinormal modes of massless scalar and Dirac fields are given in tables \ref{tableI}-\ref{tableIV}.  There one can see that the cosmological constant strongly (up to one order) suppresses the oscillation frequency and damping rate. The modes are found by the 6th order WKB formula using the Padé approximants and by the usual 3d order WKB formula. The WKB method converges only asymptotically, so that one is guaranteed that each next order will be more accurate than the previous. However, in a great number of examples \cite{Macedo:2024dqb,Bolokhov:2023ruj,Konoplya:2019hlu} we observe that if the difference between two WKB orders, say, 4th and 6th or 5th and 7th is relatively small, we have a kind plateau, where the WKB results are sufficiently accurate. The Padé approximants obey the same logic: if the frequency produced at some WKB order with Padé approximants chosen with different splitting given by $\tilde{m}$ lead to close values of frequencies, the WKB method is stable and the obtained results are reliable. Here we will use 6th WKB order with  $\tilde{m}=4$.
which proved to be optimal in the majority of cases \cite{Konoplya:2019hlu}. For comparison we will also use the usual 6th order (and 3d) order WKB formulas, using the relative error:
\begin{equation}
E = \Biggr|\frac{\omega_{6th} - \omega_{6th_{Padé}}}{\omega_{6th}}\Biggr|.
\end{equation}

Indeed, comparison with time domain integration shows that the 6th order WKB formula with $\tilde{m}=4$ Padé approximants is sufficiently accurate, providing a relative error of less than one percent for $\ell \geq 1$. For $\ell=0$ we also use the time-domain integration method which shows that at asymptotically late times purely imaginary, i.e. non-oscillatory, frequencies dominate in a signal. Comparison with the 6th order WKB data given in tables \ref{tableI}-\ref{tableIV}  demonstrates reasonable accuracy of the 6th order WKB method with Padé approximants at $\ell =0$ and very good accuracy at all higher multipoles.

From tables  \ref{tableI}-\ref{tableIV} we can see that the oscillation frequency and damping rate decrease strongly when the charge $Q$ is increased for massless scalar and Dirac field. The difference between 6th WKB order with and without Padé approximants  is less than $1\%$ for $\ell \geq 1$ , but may exceed $100\%$ for near extreme black holes at $\ell=0$ scalar perturbations. In the latter case we can see that the WKB data with Padé approximants does not give sufficiently accurate results. Moreover, the purely exponential, non-oscillatory mode, dominating the asymptotic decay cannot be reproduced within the WKB approach in principle.  Therefore, time-domain integration is desirable for checking this case. Comparison with the time-domain integration shows that the difference with the 6th order WKB method with Padé approximants is usually considerably smaller than one percent. For example, for Dirac perturbations at $Q=0.5$ in table \ref{tableIII} we have $\omega M = 0.084166 - 0.040877 i$, while the time-domain integration gives  $\omega M = 0.0843141 - 0.0408464 i$. 

From table \ref{tableVII} we see that the coupling constant $a$ softly suppresses the real and imaginary parts of $\omega$ and the latter is more sensitive to the change of $a$ than the real oscillation frequency given by $Re (\omega)$. Therefore the ratio $Re (\omega)/|Im (\omega)|$, which is proportional to the quality factor, increases when $a$ is increased. This means that the charged black hole proves to be a better oscillator when the correction from the non-linear electrodynamics is taken into account.  At the same time  from table \ref{tableVII} and example of comparison with the time-domain integration given in fig. \ref{fig8}, we can see that the the time-domain integration gives $\omega = 0.199692 - 0.0645261 i$, while the WKB formula with Padé approximants $\omega = 0.199706 - 0.064560 i$ relative error for the Schwarzschild branch of modes is considerably smaller than $0.05 \%$, while the overall effect due to the non-zero coupling $a$ is about $5 \%$.  Therefore, we conclude that the error is at least a couple of orders smaller than the observed effect for $\ell \geq 1$ and we can still rely upon the 6th order WKB method with Padé approimants.

\subsection{Massive case}

When the mass of the field is turned on, we can distinguish three regimes determined by small, large and intermediate values of $\mu M$.  As can be seen in fig. x when $\mu M \gg 1$ the asymptotic decay is exponential and oscillatory, reflecting the fact that in this regime the modes of pure..

In the limit of a vanishing black hole mass $M \rightarrow 0$, the exact analytic expression for quasinormal modes of the pure de Sitter space is known for the scalar field \cite{Lopez-Ortega:2012xvr, Lopez-Ortega:2007vlo}:
\begin{equation}\label{exact1}
i \omega_{n} R = \ell + 2n + \frac{3}{2} \pm \sqrt{\frac{9}{4} - \mu^2 R^2},
\end{equation}
for
\begin{equation}
\frac{9}{4} > \mu^2 R^2,
\end{equation}
and as follows:
\begin{equation}\label{exact2}
i \omega_{n} R = \ell + 2n + \frac{3}{2} \pm i \sqrt{\mu^2 R^2-\frac{9}{4}},
\end{equation}
for
\begin{equation}
\frac{9}{4} < \mu^2 R^2.
\end{equation}
Here, $R=\sqrt{3/\Lambda}$ is the de Sitter radius.

In the regime of small and zero $\mu M$, the quasinormal modes of the pure de Sitter space are purely imaginary, i.e., exponentially decaying and non-oscillatory,  dominating the asymptotic decay. This agrees with observations in  \cite{Konoplya:2022xid} and \cite{Konoplya:2024ptj}.

In the near extreme regime, when the cosmological horizon approach the event horizon, the quasinormal modes approach zero (see fig. 4), which is in concordance with exact solutions obtained via Poschl-Teller approximation and its generalizations \cite{Cardoso:2003sw,Churilova:2021nnc}.

On the example presented in table \ref{tableVIII} we see that once $\mu M$ is not very small, the WKB data are reliable even for $\ell=0$ perturbations, because the difference between the ordinary WKB results and those with Padé approximants is tiny. This is confirmed also by the time-domain integration data presented in figs. 4 and 5.  The non-zero coupling $a$ leads to slight decreasing of the real oscillation frequency and increasing of the damping rate.

\section{The Reissner-Nordström limit}

In the regime  $\mu M \gg 1$,  quasinormal modes that dominate during the asymptotic decay can be found analytically by  higher-order WKB expansion and expansion in terms of powers of $1/\mu$ and $Q$ without resorting to the $1/\ell$ expansion as in \cite{Konoplya:2023moy}. Using the designation of \cite{Konoplya:2023moy}:
\begin{equation}
K = n + \frac{1}{2},
\end{equation}
and introducing, for compactness of the final analytic expressions, a quantity $\sigma$:
\begin{equation}
\sigma = (9 M^2 \Lambda)^{1/6},
\end{equation}
in the limit of linear electrodynamics $a=0$ we obtain the location of the peak of the scalar field effective potential

\vspace{5mm}
\begin{widetext}
\begin{equation}
r_{0} = \frac{3 M}{\sigma ^2}-\frac{Q^2}{3
   M}+\frac{\frac{\left(\sigma ^2-1\right) \left(\ell ^2+\ell
   +\sigma ^2\right)}{3 M}+\frac{\sigma ^2 \left(\left(3
   \sigma ^2-4\right) \ell ^2+\left(3 \sigma ^2-4\right)
   \ell -\sigma ^2\right) Q^2}{27
   M^3}}{\mu ^2}+O (\mu^{-3}, Q^{3}).
\end{equation}

Then, we use the obtained expression for the position of the peak of the effective potential in the 6th order WKB formula \cite{Schutz:1985km, Iyer:1986np, Konoplya:2003ii} 
\begin{equation}\label{eq:11}
   \frac{i (\omega^2 - V_{0})}{\sqrt{-2 V_{0}^{\prime \prime}}}
   - \sum_{i=2}^{i=6} \lambda_{i} =n+\frac{1}{2}.
\end{equation}
Here $\lambda_{i}$ is the $i$-th order WKB correction, $n$ is the overtone number,  $V_{0}$ is the value of the effective potential in its maximum. The WKB corrections $\lambda_{i}$ depend on derivatives of orders up to $2 i$ of the effective potential in its peak and the explicit form of the WKB corrections were obtained in  \cite{Schutz:1985km, Iyer:1986np, Konoplya:2003ii}.

Finally we obtain the quasinormal frequencies for the scalar field perturbations,
$$\omega_{n} = \mu  \left(\sqrt{1-\sigma ^2}+\frac{\sigma ^4 Q^2}{18 M^2
   \sqrt{1-\sigma
   ^2}}\right)+\left(-\frac{i K \sigma ^3
   \sqrt{1-\sigma ^2}}{3 M}-\frac{i Q^2 K \sigma ^5 \left(2
   \sigma ^2-1\right)}{54 M^3 \sqrt{1-\sigma
   ^2}}\right)-$$
   $$\frac{\sigma ^4
   \sqrt{1-\sigma ^2} \left(-72 \ell ^2-72 \ell +29 \sigma
   ^2+12 K^2 \left(\sigma ^2-1\right)-11\right)}{1296
   M^2 \mu}-$$
\begin{equation}
\frac{\left(\sigma ^6 \left(113 \sigma ^4-121 \sigma
   ^2+72 \ell ^2 \left(3 \sigma ^2-4\right)+72 \ell  \left(3
   \sigma ^2-4\right)+12 K^2 \left(\sigma ^4+\sigma
   ^2-2\right)+26\right)\right) Q^2}{23328 \mu \left(M^4
   \sqrt{1-\sigma ^2}\right)} +O(\mu^{-2}, Q^{3})).
\end{equation}
\end{widetext}

When $a \neq 0$, the resultant analytic expression for quasinormal modes is too cumbersome.

\section{Conclusions}

In the present paper we have calculated quasinormal frequencies of the massive scalar and massless Dirac fields in the background of the charged asymptotically de Sitter black holes when corrections of the Heisnebreg-Euler non-linear electrodynamics are taken into account. An important feature of the spectrum is the existence of two branches of quasinormal modes: Reissner-Nordstrom branch, corrected by the non-zero cosmological constant, and de Sitter branch, deformed by the adding a black hole. It is the latter branch that dominates in the asymptotically late times and these second branch of modes can be calculated for large and intermediate $\mu M$ with the WKB method sufficiently accurately. 

We have shown that while the usual 3d order WKB formula is not accurate enough, especially for $\ell=0$ scalar field perturbations, the 6th order formula with the Padé approximants works quite well, unless the mass of the field is very small, the value of the cosmological constant is near extreme and $\ell=0$. In that case, one should rely upon time-domain integration.

We have not considered a massive Dirac field, because for this case the effective potential will depend on $\omega$ and the whole procedure is more involved, Nevertheless, the time-domain integration and WKB approach can be applied for that case as well after some modifications. 

In the limit of vanishing coupling $a$ we produce a compact analytic expression for $\omega$ in the regime of large $\mu M$ without resorting to $1/\ell$ expansion.

\acknowledgments
The author would like to thank R. A. Konoplya for useful discussions and careful reading of the manuscript. This work was supported by RUDN University research project FSSF-2023-0003.

\end{document}